\documentclass[twocolumn,showpacs,preprintnumbers,amsmath,amssymb,aps,showkeys]{revtex4}
\usepackage{graphicx,bm}

\begin{document}

\title{On The Critical Packet Injection Rate Of A Preferential Next-Nearest
Neighbor Routing Traffic Model On Barab\'{a}si-Albert Networks}

\author{H.~F. Chau}
\email{hfchau@hkusua.hku.hk}
\author{H.~Y. Chan}
\altaffiliation{Present address: Department of Physics, University of
Houston, 617 Science \& Research Building~1, Houston, Texas 77204-5005, USA}
\author{F.~K. Chow}
\affiliation{Department of Physics, and Center of Theoretical and
Computational Physics, University of Hong Kong, Pokfulam Road, Hong Kong}
\date{\today}

\begin{abstract}
Recently, Yin \emph{et al.} [Eur. Phys. J. B 49, 205 (2006)] introduced an
efficient small-world network traffic model using preferential next-nearest 
neighbor routing strategy with the so-called path iteration avoidance (PIA)
rule to study the jamming transition of internet.
Here we study their model without PIA rule by a mean-field analysis which
carefully divides the message packets into two types. Then, we argue that
our mean-field analysis is also applicable in the presence of PIA rule
in the limit of a large number of nodes in the network. 
Our analysis gives an explicit expression of the critical packet injection
rate $R_c$ as a function of a bias parameter of the routing strategy
$\alpha$ in their model with or without PIA rule. In particular, we predict
a sudden change in $R_c$ at a certain value of $\alpha$. These predictions
agree quite well with our extensive computer simulations.
\end{abstract}


\maketitle

\section{Introduction \label{sec:intro}}
 
Complex networks with small-world property exist in many natural and social systems, such as food web, the  
internet~\cite{internet1,internet2}, the world wide web~\cite{www}, and the world-wide airport network (WAN)~\cite{wan}.
In 1999, Barab\'{a}si and Albert proposed a scale-free growing model (BA network) with a preferential attachment 
mechanism to mimic a growing small-world network in the real world~\cite{BA}. Their model 
stimulated the interest of the physics community to study complex networks by statistical physical means~\cite{BA_review,Newman_review}.
One of the goals of these studies is to understand the dynamical processes taking place behind the underlying structure. 

It is instructive to study the traffic capacity of a network. We may start
by considering the simple-minded situation in which message packets are
injected randomly into the nodes of the network at a fixed rate. Each
packet has a randomly assigned destination node. And each node in the
network has a finite message-forwarding rate. Clearly, an important factor
affecting network traffic capacity is the routing strategy, namely, how
each node forwards its out-going message packets to its nearest neighbors.
The performance indicator is the maximum free-flowing traffic capacity
characterized by the critical packet generation rate $R_{c}$. More
precisely, $R_{c}$ is the supremum number of new packets that can be
injected into the network per unit time step without causing
congestion~\cite{congestion1,congestion2}. Here, congestion means that
the average rate of change of the number of packets in some node is positive.
(Actually, this performance indicator is not overly stringent for the
model investigated in this paper as we find that the number of packet in
almost all nodes steadily increases over time without saturation whenever
$R > R_c$.) The more efficient the routing strategy, the larger the value
of $R_{c}$. For a sufficiently large random network, the routing strategy
cannot depend on the network topology because this information is not
available to each node.
Thus, it is reasonable to confine ourselves to study local routing
strategies.

Perhaps the simplest local routing strategies are the ones that use
information on the nearest neighbors of individual node~\cite{Adamic,Tadic}.
Recently, based on these nearest-neighbors-based strategies, a new routing
strategy called the preferential next-nearest-neighbor (PNNN) searching
strategy was proposed by Yin \emph{et al.}~\cite{Yin} in which the
performance is better than those using nearest neighbor routing. As the name
suggests, in PNNN, a message packet looks for its destination among the next
nearest neighbors of the node it currently stays. 
If the destination cannot be found in this way, the message packet will be
forwarded to a neighboring node by a biased
random walk with a preferential probability which depends on a parameter 
called preferential delivering exponent $\alpha$.
To speed up packet delivery, Yin \emph{et al.} added in their routing strategy
the path iteration
avoidance (PIA) rule, which states that a packet cannot travel through an edge
more than twice.

As a model of scale-free network traffic with potential applications in the
internet and the world wide web, the use of PIA rule is problematic.
A message packet, unlike a human driver, cannot automatically remember the
path it has traveled. This additional piece of information, whose length
grows linearly with the time since creation of the packet, may either be
stored in, say, a central registry, or
attached to the message packet itself. Thus, the cost of inquiring this
information from the registry or transmitting it through an edge alongside with
the message packet cannot be ignored. Furthermore, additional computational
cost, which also scales linearly with the time since the creation of the
packet, is needed for a
node to process this historical path information in accordance with the PIA
rule. All these factors make the effective message packet forwarding rate
a function of the time since the packet creation.
Unfortunately, the PNNN routing strategy of Yin \emph{et al.}~\cite{Yin} does
not take these extra communication and computational costs into account. This
is why we believe that PIA rule is not very realistic.

In Sec.~\ref{sec:model}, we briefly review the network traffic model proposed
by Yin \emph{et al.}~\cite{Yin} using the PNNN strategy. Then we perform
mean-field analytical calculation for the dynamics of their model with and
without the PIA rule in Sec.~\ref{sec:cal}. In both cases, we find an abrupt
change in the dependence of $R_c$ on $\alpha$ at certain value of $\alpha$.
We also give the physical reason behind such change. 
In Sec.~\ref{sec:sim}, we compare the mean-field calculations with our
extensive numerical simulation results of $R_{c}$ against $\alpha$. We also
show in this Section that the network size used in Yin \emph{et al.}'s
numerical simulations is not large enough to reveal the thermodynamic behavior
of their model. Finally, we give a brief summary and discuss the effectiveness
of the PNNN strategy in Sec.~\ref{sec:sum}. 

\section{The PNNN+PIA And PNNN-PIA Models \label{sec:model}}

Yin \emph{et al.} proposed and studied the following network traffic model on
a BA network~\cite{Yin}. (Here we call their model with and without the PIA
rule PNNN+PIA and PNNN-PIA, respectively.)
Their model consists of a random but fixed BA network with $N$ nodes.
We denote the set of all nodes in this network by ${\mathbb V}$.
We further denote the degree of the node having the least (greatest) number of
nearest neighbors in the network by $k_{\min}$ ($k_{\max}$). That is to say,
\begin{equation}
k_{\min} \equiv \min_{i\in {\mathbb V}} k_i \label{k_min_def}
\end{equation}
and
\begin{equation}
k_{\max} \equiv \max_{i\in {\mathbb V}} k_i , \label{k_max_def}
\end{equation}
where $k_i$ is the degree of the node~$i$. Recall that
BA network is generated by connecting each newly added node to
$m$ existing nodes in a careful way~\cite{BA}. Hence,
\begin{equation}
k_{\min} = m . \label{k_min_equal_m}
\end{equation}
Further recall that during the generation of a BA network,
the average degree of the node added to the network $t_\text{elapse}$ ago
equals $m (N / t_\text{elapse})^{1/2}$~\cite{BA}. Thus,
\begin{equation}
k_{\max} \approx  m \sqrt{N} . \label{k_max_estimate}
\end{equation}
We denote the adjacency matrix of the network by $A$. That is, $A_{ij} = 1 (0)$ if
there is an (no) edge between nodes~$i$ and~$j$.

In PNNN+PIA and PNNN-PIA,
each node has an unlimited buffer, known as load, to store packets. At each time step, each of the $R$ packets 
is added to a randomly chosen source node of the network with a randomly chosen destination node. Note that
simulations reported by Yin \emph{et al.} in Ref.~\cite{Yin} were performed by considering only integer values of $R$. 
In contrast, we allow a real-valued $R$. More precisely, we inject a
message packet into a node with probability $R/N$ in each time step. 

Each node can send out at most 
$C\geq 1$ packets to its nearest neighbors using the first-in-first-out rule.
That is to say, packets entering a node first will be sent out first. 
Each out-going packet first searches through all the next nearest neighbors of the node to which it currently belongs. 
If its destination is located in this search,
the packet will be forwarded to one of the neighbors connecting the destination 
and the current node. And in the next time step, this packet will be
forwarded to the destination and then removed from the network.
If the destination of an out-going message packet cannot be found in such a
search, it will be randomly forwarded from its current node (say, node~$i$)
to one of the neighbors (say, node~$j$) with probability
\begin{equation}
\Pi_{ij} = \frac{k_{j}^{\alpha}}{\sum_{\ell\in {\mathbb V}} A_{i\ell}
k_{\ell}^{\alpha}}, \label{free_prob}
\end{equation}
where $\alpha$ is a fixed parameter known as the preferential delivering
exponent. Note that the sum in the above equation
can be regarded as a restricted sum over
the nearest neighbors of the packet's current node~$i$.

The only difference between PNNN+PIA and PNNN-PIA is that PIA rule is
present in the former model while absent in the latter.
Recall that PIA rule demands each packet to travel through the same edge at most
twice~\cite{Yin}. In the event
that a message packet has nowhere to go due to the PIA rule, the packet will
be removed from the network. And for $\alpha \in [-4,2]$,
only a very small percentage
of packets are removed from the network in this way~\cite{Wang_private}.

Clearly, historical path information of a packet is needed to decide where it
will go in the next time step with the adoption of PIA rule.
As we have mentioned in Sec.~\ref{sec:intro},
extra communication and processing costs are required to forward a
message packet together with its historical path information in the network to
its neighboring node. Thus, it is less efficient to forward an old packet than
a newly created one.
In this respect, PIA rule is not consistent with the rule that the
message forwarding capability of a node is independent of
the age of the forwarding packets. This is a serious problem
because Yin \emph{et al.} found by numerical simulation that the packet
lifetime, which is the time between its injection and removal, roughly obeys a
power law distribution~\cite{Yin}.

Although Yin \emph{et al.} has briefly studied the PNNN-PIA model numerically in
Ref.~\cite{Yin}, their focus was on the PNNN+PIA model. They
found that the critical packet generation rate $R_c$ is increased by adopting
the PIA rule.
More importantly, using numerical simulation up to $N \approx 5000$ with
$R$ restricted to integers only,
they found that $R_c$ is a decreasing function of
$\alpha$ for the PNNN+PIA model. In addition, based on their
simulations in the range $\alpha \in [-4,2]$, they believed that for a fixed
$N$, the value of $R_c$ is a constant whenever $\alpha\leq -2$~\cite{Yin}.

An interesting common feature of the PNNN+PIA and PNNN-PIA models is that
as long as there
are no more than $C$ message packets staying in a node at any time,
the message packets behave like independent particles in the sense that
their motions in the BA network are independent of each other.
This property is important in our subsequent discussions.

\section{Mean-field Analysis \label{sec:cal}}

\subsection{The PNNN-PIA Model \label{subsec:no_PIA}}
We try to calculate the $R_c$ against $\alpha$ curve for the PNNN-PIA model
by mean-field approximation. The validity of the approximations made in our
calculation will be discussed and justified in Sec.~\ref{sec:sim}. Let us
begin by classifying the packets into two types. A packet is called a
destination located packet (DLP) if it has successfully found a path to its
destination. By the rules of PNNN-PIA, the destination of a DLP must be one
of the nearest or next nearest neighboring nodes of its current location.
Otherwise, the packet is known as a destination seeking packet (DSP). Since a
newly injected packet has not found out its path to the destination yet, it
must a DSP. A DSP moves randomly to
its neighboring node with probability given by Eq.~(\ref{free_prob}). We
denote the numbers of DLPs and DSPs in node~$i$ at time $t$ by $n_{l,i}(t)$
and $n_{s,i}(t)$, respectively. 

We say that a network is in free-flow state if each node, on average, can
forward all its loads in the next time step.  (In other words, the average load
of each node is at most $C$ in each time step.)
In this case, node~$i$ can, on average, send out all its $n_{s,i}(t)$
message packets at any time $t$.
At the same time, node~$i$ receives, on average, $R/N$ DSPs by
packet generation.
Since the number of 4-cycles in a BA network scales like $\left[ m \log (N) /2
\right]^4 / 4$~\cite{BA_4-cycle}, the probability that two next
nearest neighboring nodes are connected to more than one common node goes to
0 in the large $N$ limit. So
the number of next nearest neighbors for node~$i$ is approximately equal to
$\sum_{j\in {\mathbb V}} A_{ij} (k_j - 1)$ in the large $N$ limit.
Consider a DSP that reaches the node~$i$ for the first time. Then, the
probability $\Phi_i$ that it can locate a path to its destination
in the next time step is given by
\begin{equation}
 \Phi_i \approx \frac{1}{N} \sum_{j\in {\mathbb V}} A_{ij} (k_j - 1) . \label{Phi_i}
\end{equation}
In contrast, suppose the DSP has reached the node~$i$ more than once, then it
has no chance to find the path to its destination in the next time step as the
next nearest neighbors of node~$i$ has been searched during its
previous visit to node~$i$.
Let $\lambda_j$ be the average number of visit of a DSP to node~$j$ given that
it has visited node~$j$ at least once. Then,
using the mean-field approximation similar to that used in
Ref.~\cite{mean_field}, the number of DSP in free-flow state satisfies
\begin{equation}
\frac{dn_{s,i}(t)}{dt} \approx \frac{R}{N} - n_{s,i}(t) +
 \sum_{j\in {\mathbb V}} A_{ij} n_{s,j}(t) \Pi_{ji} \left(
 1-\frac{\Phi_j}{\lambda_j} \right) . \label{free_eqn_t}
\end{equation}
Note that $1 - \Phi_j/\lambda_j$ is the probability
that a DSP at node~$j$ will not change to a DSP in the next time step.
Thus, the last term in the R.H.S. of the above equation
is the average number of DSPs received by node~$i$ from its neighbors.

We want to study the equilibrated distribution of DSPs for a
typical node in the free-flow state as a function of the degree
of the node. And we do so by investigating
\begin{equation}
 n_s(k) \equiv \frac{\left\langle \sum_{i\in {\mathbb V}} \delta_{k_i,k} n_{s,i}(t)
 \right\rangle_t}{\sum_{i\in {\mathbb V}} \delta_{k_i,k}} , \label{n_s_k_def}
\end{equation}
where $\delta_{k_i,k}$ is the Kronecker delta and $\left\langle \cdots \right\rangle_t$
represents the time average of its argument.  Upon equilibration,
\begin{equation}
 \left\langle n_{s,i}(t) \right\rangle_t \approx \frac{R}{N} +
 \sum_{j\in {\mathbb V}} A_{ij} \left\langle n_{s,j}(t) \right\rangle_t
 \Pi_{ji} \left( 1 - \frac{\Phi_j}{\lambda_j} \right) . \label{in-out-balance}
\end{equation}
Although BA network does not show assortative
mixing~\cite{no-assortative_mixing}, it exhibits non-trivial but weak
degree-degree correlation between neighboring nodes~\cite{BA_review}.
Combined with the fact that the trajectory of a DSP is history independent for
PNNN-PIA, it makes sense to ignore this degree-degree correlation in our
mean-field analysis. By ignoring this correlation, we know that
for any function $f(k)$,
\begin{equation}
\sum_{\ell\in {\mathbb V}} A_{j\ell} f(k_{\ell}) \approx k_{j} {\mathcal D}
\label{no_assort}
\end{equation}
where ${\mathcal D}$ is a functional of $f$.
Most importantly, ${\mathcal D}$ is independent of $k_{j}$.
As a result, using Eqs.~(\ref{free_prob})--(\ref{Phi_i}),
we can re-express Eq.~(\ref{n_s_k_def}) as
\begin{widetext}
\begin{equation}
n_s(k) \approx \frac{R}{N} + \frac{k^\alpha}{\sum_{i\in {\mathbb V}}
\delta_{k_i,k}} \sum_{i,j\in {\mathbb V}} \left\{
\frac{\delta_{k_i,k} A_{ij} n_{s}(k_j)}{\sum_{\ell\in {\mathbb V}} A_{j\ell}
k_\ell^\alpha} \left[ 1 - \frac{\sum_{\ell\in {\mathbb V}} A_{j\ell}
(k_\ell - 1)}{\lambda_j N} \right] \right\}
\approx \frac{R}{N} + D k^{\alpha + 1} \sim D k^{\alpha+1}
\label{n_s_k_i_form} 
\end{equation}
\end{widetext}
for some $D>0$ independent of $k$. Of course, $D$ depends on $\alpha$ and
$N$.

To derive the mean field equation for $n_{l,i}(t)$ in the free-flow state,
we consider a DSP currently located at $j$, which is a neighboring node of $i$.
Suppose this is the first time for this DSP to visit a neighboring node of
$i$.  Then, on average, the chance for
this packet to turn into a DLP and then forwarded to node~$i$
in the next time step equals $\Phi_j [ (k_i - 1) / \sum_\ell A_{j\ell} (k_\ell - 1) ] = (k_i - 1) / N$
in the large $N$ limit.
In contrast, if this is not the first time for the DSP to visit a neighboring
node of $i$, then it has no chance to be forwarded to node~$i$ as a DLP in the
next time step.
This is because the DSP should have converted into a DLP after its first visit
to a neighboring node of $i$.
Suppose a DSP was located at a neighboring node of $i$ at time step $t-1$.
Suppose further that this packet is forwarded to node~$i$ at time step $t$.
Then it must be found in a neighboring node of $i$ at time step $t+1$.
In contrast, suppose the packet is forwarded to a node other than $i$ at
time step $t$, then the chance that it will come back to
a neighboring site of $i$ will be roughly proportional to $k_i$.
Hence, the average number of times for a DSP to visit neighboring nodes of
$i$ given that it has visited a neighboring node of $i$ once equals
$(\mu + \nu k_i)$ for some $\mu > 1$ and $\nu > 0$ independent of $k_i$.

It is obvious that $\mu$ is independent of $N$. In what follows, we argue
that $\nu$ is also independent of $N$. BA network is a small world network
without showing any assortative mixing~\cite{no-assortative_mixing}. And 
the packet forwarding rule in Eq.~(\ref{free_prob}) does not depend on the
historical path of the packet. So, message packets are essentially
performing random walk in the network in the first few steps after its
injection. Consequently, the probability distribution of the first return
time of a random walker should scale like $t^{-\xi}$ for some
$\xi > 0$ and sufficiently small $t$~\cite{Redner_book}.  Moreover, $\xi$ is
independent of $N$.
On the other hand, when $t$ is approximately greater than
the average square distance between two nodes in the network
$\left\langle d^2 \right\rangle$, finite size effect of the network will
affect the probability distribution of the first return time of a random
walker so that the $t^{-\xi}$ scaling will no longer be valid.
Indeed, this is what
Almaas \emph{et al.} have found in their numerical study of the first return
time for random walk in a certain small world network. More importantly, they
found that the probability distribution of the first return time collapses
to a single scaling relation by rescaling both the first return time $t$
and the probability $P$ by $\left\langle d^2 \right\rangle$~\cite{Almaas}.
Since the packet lifetime $\tau$ scales roughly as
$\left\langle d^2 \right\rangle$, we conclude that
$\nu$ is independent of
$N$. Nevertheless, both $\mu$ and $\nu$ are functions of $\alpha$. But the
form of Eq.~(\ref{free_prob}) assures that $\mu$ and $\nu$ are not
sensitively dependent on $\alpha$ in the sense that $\mu$ and $\nu$ scale
polynomially instead of, say, exponentially with $\alpha$.

Utilizing all these information, we may write the mean field equation for
$n_{l,i}(t)$ in free-flow state as follow:
\begin{equation}
\frac{dn_{l,i}(t)}{dt} \approx
-n_{l,i}(t) + \sum_{j\in {\mathbb V}} \frac{A_{ij} n_{s,j}(t) (k_i - 1)}{N
(\mu+\nu k_i)} . \label{dir_eqn_t}
\end{equation}

Note that the average number of DLPs for a typical degree $k$ node in the
free-flow state upon equilibration:
\begin{equation}
 n_l(k) \equiv \frac{\left\langle \sum_{i\in {\mathbb V}} \delta_{k_i,k}
 n_{l,i}(t)
 \right\rangle_t}{\sum_{i\in {\mathbb V}} \delta_{k_i,k}} \label{n_l_k_def}
\end{equation}
satisfies
\begin{equation}
 n_l(k) \sum_{i\in {\mathbb V}} \delta_{k_i,k} \approx \frac{k - 1}{N
 (\mu+\nu k)} \sum_{i,j\in {\mathbb V}} \delta_{k_i,k} A_{ij} n_s(k_j) .
 \label{n_l_k_intermediate}
\end{equation}
By ignoring the degree-degree correlation between neighboring nodes as in the
derivation of the scaling relation for $n_{s}(k)$, we have
\begin{eqnarray}
n_{l}(k) & \approx & \frac{\left\langle n_s(k_i) \right\rangle_{i\in{\mathbb V}}
k (k-1)}{N (\mu+\nu k)} \label{dir_dis1} \\
& \sim & \frac{\left\langle n_s(k_i) \right\rangle_{i\in {\mathbb V}} k}{N\nu} .
\label{dir_dis}
\end{eqnarray}
As we shall see in Sec.~\ref{sec:sim}, the value of $\nu$ is of order of
0.01 for most values of $\alpha$.
Thus, for network size $N \lesssim 5000$ such as those used in the
simulations reported in Ref.~\cite{Yin}, $n_{l}(k)$ varies quadratically
rather than linearly in most of the domain $[k_{\min},k_{\max}]$.
In this respect, Yin \emph{et al.}'s numerical results did not reflect the
properties of the system in the large $N$ limit. We shall discuss more along
this line in Sec.~\ref{sec:sim}.

Upon equilibration, the average number of packet residing on a typical
degree~$k$ node equals
\begin{eqnarray}
 n(k) & \equiv & n_s (k) + n_l (k) \nonumber \\
 & \approx & \frac{R}{N} + D k^{\alpha+1} + \frac{\left\langle n_s (k_i)
 \right\rangle_{i\in\mathbb V} k(k-1)}{N(\mu+\nu k)} \label{n_full_def} \\
 & \approx & \frac{R}{N} + D k^{\alpha+1} +
 \frac{\left\langle n_s (k_i) \right\rangle_{i\in\mathbb V} k}{N\nu}
 \label{n_def}
\end{eqnarray}
in the large $N$ limit.

\subsubsection{A Simplifying Assumption \label{subsub:ideal_case}}
In this Subsection, we make the simplifying assumption that the expressions of
$n_s(k)$ and $n_l(k)$ in Eqs.~(\ref{n_s_k_i_form}) and~(\ref{dir_dis1})
are exact throughout the entire domain $[k_{\min},k_{\max}]$.
Then, it is clear that Eq.~(\ref{n_full_def}) is an
increasing function of $k$ for $\alpha > -1$.
Hence, the maximum value for the last line of Eq.~(\ref{n_def}) in this domain
is attained when $k = k_{\max}$.
And in the case of $\alpha < -1$, 
Eq.~(\ref{n_full_def}) is a continuous
function with one local minimum point in the interval
$[k_{\min},k_{\max}]$. So, again in this interval, $n(k)$
attains its maximum value at the boundary.
To find out the exact location at which the maximum value is attained,
we have to find an
expression for $\left\langle n_s(k_i) \right\rangle_{i\in {\mathbb V}}$ first.

According to Albert and Barab\'{a}si, the probability distribution of nodes of
degree $k$ for a BA network is given by
\begin{equation}
p(k) \sim E k^{-\gamma} , \label{k_dis}
\end{equation}
where $E$ is the normalization constant and $\gamma = 3$~\cite{BA_review}.
So the normalization constant $E$ can be rewritten as
\begin{equation}
 E = \left( \int_{k_{\min}}^{k_{\max}} k^{-\gamma} dk \right)^{-1} \approx
 (\gamma-1) m^{\gamma-1} . \label{E_value}
\end{equation}
Using our assumption that Eq.~(\ref{n_s_k_i_form}) is valid over the entire
interval $[k_{\min},k_{\max}]$, we arrive at
\begin{widetext}
\begin{equation}
\left\langle n_s(k_i) \right\rangle_{i\in {\mathbb V}} =
\int_{k_{\min}}^{k_{\max}} p(k) n_s(k) dk
\approx \frac{D(\gamma - 1) m^{\gamma-1} \left(
k_{\max}^{\alpha-\gamma+2} - k_{\min}^{\alpha-\gamma+2} \right)}{\alpha
-\gamma+2}
\label{ensemble_average_n_s}
\end{equation}
in the large $N$ limit provided that $\alpha \neq \gamma-2 = 1$.

By substituting Eqs.~(\ref{k_min_equal_m}), (\ref{k_max_estimate})
and~(\ref{ensemble_average_n_s}) into Eq.~(\ref{n_def}) together with the fact
that $\nu$ is independent of $N$ and is not sensitively
dependent on $\alpha$, we find
\begin{equation}
 n(k_{\min}) - n(k_{\max}) \approx D m^{\alpha+1} \left\{
 1 - N^{(\alpha+1)/2} + \frac{2m \left[ 1-N^{(\alpha-1)/2} \right] \left(
 1-N^{1/2} \right)}{N\nu (1-\alpha)} \right\} > 0 \label{n_k_min_vs_n_k_max}
\end{equation}
\end{widetext}
in the large $N$ limit whenever $\alpha < -1$. Thus, the maximum of $n(k)$ is
attained at $k = k_{\min}$ provided that $\alpha < -1$ and $N\rightarrow\infty$.

To summarize, the maximum value of $n(k)$ is always attained either at
$k = k_{\min}$ or $k = k_{\max}$. By denoting the value $k$ at which $n(k)$
reaches its maximum value by $k_c$, we have
\begin{equation}
\lim_{N\rightarrow\infty} k_c = \left\{ \begin{array}{ll}
k_{\max} & \mbox{if~} \alpha > -1 , \\
k_{\min} & \mbox{otherwise.}
\end{array} \right.
\label{k_c_def}
\end{equation}
And from Eq.~(\ref{n_k_min_vs_n_k_max}), for any fixed $N > 0$, there is a
critical value of $\alpha = \alpha_c \leq -1$ above (below) which $k_c =
k_{\max}$ ($k_c = k_{\min}$). Besides,
\begin{equation}
\lim_{N\rightarrow\infty} \alpha_c = -1 . \label{alpha_c}
\end{equation}

There is an important consequence of the above findings. By gradually
increasing the packet injection rate $R$, the first congested node must be
the one with the largest value of $n(k)$. Therefore, the critical packet
injection rate $R_c$ is reached when congestion occurs at a smallest
(largest) degree node whenever $\alpha < \alpha_c$ ($\alpha > \alpha_c$).
This change in the type of node that congests first upon a gradual increase
in $R$ results in the discontinuity of $R_c$ at
$\alpha = \alpha_c$. Actually, $R_c$ attains its maximum value at
$\alpha = \alpha_c$ in the large $N$ limit. To see why, we consider the
situation when the network is at its maximal capacity. In this situation,
$R=R_{c}$ and the maximum number of packets in some nodes should be $C$.
From Eq.~(\ref{n_full_def}), $R_c$ satisfies
\begin{eqnarray}
C & \approx & \frac{D (\gamma-1) m^{\gamma-1} \left(
k_{\max}^{\alpha-\gamma+2} - k_{\min}^{\alpha-\gamma+2} \right) k_c (k_c -1)}{N
(\alpha-\gamma+2)(\mu+\nu k_c)} + \nonumber \\
& & \,\frac{R_c}{N} + D k_c^{\alpha+1} . \label{kc_sup}
\end{eqnarray}
We need to eliminate $D$ in order to simplify the above equation.
We proceed by considering the average number of packets reaching their
destinations in each time step at equilibrium. This number is equal to the
average number of packets injected into the system at each time step.
Therefore,
\begin{equation}
R = \int_{k_{\min}}^{k_{\max}} N p(k) n_{l}(k)dk . \label{R_int}
\end{equation}
From Eqs.~(\ref{k_min_equal_m})--(\ref{k_max_estimate}), (\ref{dir_dis1}) and
(\ref{k_dis})--(\ref{ensemble_average_n_s}), we know that the critical
packet injection rate equals
\begin{widetext}
\begin{eqnarray}
R_c & \approx & E \left\langle n_s(k_i) \right\rangle_{i\in {\mathbb V}}
\int_{k_{\min}}^{k_{\max}} \frac{k-1}{k^2 (\mu+\nu k)} \,dk
\approx E \left\langle n_s(k_i) \right\rangle_{i\in {\mathbb V}} \left\{
\frac{\mu+\nu}{\mu^2}
\ln \left[ \frac{(\mu+\nu m) N^{1/2}}{\mu+\nu m N^{1/2}} \right] +
\frac{1}{\mu m N^{1/2}} - \frac{1}{\mu m} \right\} \nonumber \\
& \approx & \frac{D (\gamma -1)^2 m^{\alpha+\gamma} \left[ N^{(\alpha-\gamma+2)
/2} - 1 \right] \left\{ \frac{\mu+\nu}{\mu^2} \ln \left[ \frac{(\mu+\nu
m)N^{1/2}}{\mu + \nu m N^{1/2}} \right] + \frac{1}{\mu m N^{1/2}} -
\frac{1}{\mu m} \right\}}{\alpha - \gamma + 2}
\label{R_c_intermediate}
\end{eqnarray}
in the large $N$ limit provided that $\alpha \neq \gamma - 2 = 1$.

By using Eq.~(\ref{R_c_intermediate}) to eliminate $D$ in Eq.~(\ref{kc_sup}),
we find that a sufficiently large $N$ and $\alpha \neq \gamma - 2 = 1$,
\begin{eqnarray}
R_c & \approx & \frac{C N (\gamma - 1)^2 m^{\alpha+\gamma} \left(
N^{\frac{\alpha-\gamma+2}{2}} - 1 \right) \Xi}{(\gamma-1)^2 m^{\alpha+\gamma}
\left( N^{\frac{\alpha-\gamma+2}{2}} - 1 \right) \Xi + (\alpha - \gamma + 2) N
k_c^{\alpha+1} + (\gamma-1) m^{\alpha+1} \left(
N^{\frac{\alpha-\gamma+2}{2}} - 1 \right) \frac{k_c (k_c - 1)}{\mu+\nu k_c}}
\nonumber \\
& = & \min_{k\in \{ m,m\sqrt{N} \}} \!\!\left\{
\frac{C N (\gamma - 1)^2 m^{\alpha+\gamma} \left( N^{\frac{\alpha-
\gamma+2}{2}} - 1 \right) \Xi}{(\gamma-1)^2 m^{\alpha+\gamma} \left(
N^{\frac{\alpha-\gamma+2}{2}} - 1 \right) \Xi + (\alpha - \gamma + 2) N
k^{\alpha+1} + (\gamma-1) m^{\alpha+1} \left(
N^{\frac{\alpha-\gamma+2}{2}} - 1 \right) \frac{k (k - 1)}{\mu+\nu k}}
\right\} ,
\label{Rc_full}
\end{eqnarray}
\end{widetext}
where
\begin{equation}
\Xi \equiv \frac{\mu+\nu}{\mu^2} \ln \left[ \frac{(\mu+\nu m)\sqrt{N}}{\mu+\nu
m\sqrt{N}} \right] + \frac{1}{\mu m \sqrt{N}} - \frac{1}{\mu m} .
\label{Xi_def}
\end{equation}
The above equation is not only valid for the generic case. It is
straight-forward to go through the same derivation to show that
Eq.~(\ref{Rc_full}) is also valid for the singular case of $\alpha = \gamma
- 2$ as long as we take the limit $\alpha \rightarrow \gamma - 2$
rather than simply substituting $\alpha = \gamma - 2$ into
Eq.~(\ref{Rc_full}).

Although the functional forms of $\mu$ and $\nu$ are not easy to determine,
the facts that they are independent of $N$ and are not sensitively dependent on
$\alpha$ are already sufficient for us to make the following remark on the
general trend of $R_c$: For
sufficiently large $N$, $R_c$ is an increasing (decreasing) function whenever
$\alpha < \alpha_c$ ($\alpha > \alpha_c$). Besides,
\begin{widetext}
\begin{subequations}
\begin{equation}
\lim_{N\rightarrow\infty} R_c = \frac{C (\gamma - 1)^2 m^{\gamma - 1}
\left[ \frac{\mu+\nu}{\mu^2} \ln \left( \frac{\mu+\nu m}{\nu m} \right) -
\frac{1}{\mu m} \right]}{\gamma - \alpha -2} > 0
\enspace\enspace\mbox{for~} \alpha < -1 ,
\label{Rc_limit1}
\end{equation}
\begin{equation}
\lim_{N\rightarrow\infty} R_c = \lim_{N\rightarrow\infty} \frac{C (\gamma -
1)^2 m^{\gamma - 1} \left[ \frac{\mu+\nu}{\mu^2} \ln \left( \frac{\mu+\nu
m}{\nu m} \right) - \frac{1}{\mu m} \right]}{(\gamma - \alpha - 2)
N^{(\alpha + 1)/2}} = 0
\enspace\enspace\mbox{for~} -1 < \alpha < 1 ,
\label{Rc_limit2}
\end{equation}
and
\begin{equation}
\lim_{N\rightarrow\infty} R_c = \lim_{N\rightarrow\infty} \frac{C (\gamma -
1)^2 m^{\gamma - 1} \left[ \frac{\mu+\nu}{\mu^2} \ln \left( \frac{\mu+\nu
m}{\nu m} \right) - \frac{1}{\mu m} \right]}{(\alpha - \gamma + 2) N} = 0
\enspace\enspace \mbox{for~} \alpha > 1 .
\label{Rc_limit3}
\end{equation}
\end{subequations}
\end{widetext}
In other words, in the thermodynamic limit, the change in the type of nodes
that is congested first results in the maximum point of the $\alpha - R_c$
curve at $\alpha = \alpha_c$.

We may understand the occurrence of this maximum point as follows.
Clearly, there are much more small degree nodes than large degree ones in a BA
network. Since all nodes of different degree have the same message-forwarding
capability, one may attempt to increase $R_c$ by preferentially forwarding the
DSPs to small degree nodes by setting $\alpha < 0$. If $\alpha_c < \alpha < 0$,
the bias towards sending DSPs to small degree nodes is not yet sufficient.
Hence, jamming at $R = R_c$ occurs in the largest degree node because too many
DLPs move to this node per unit time step. In contrast, if $\alpha <
\alpha_c$, the bias towards sending DSPs to small degree nodes is too strong
that the smallest degree nodes are jammed by the influx of DSPs. In this
respect, it is not surprising for our mean-field calculations to find that
$R_c$ is an increasing (decreasing) function of $\alpha < \alpha_c$ ($\alpha
> \alpha_c$).

They may break down near $k_{\min}$ and $k_{\max}$. As a result, the expression
for $D$ in Eq.~(\ref{R_c_intermediate}) should only be regarded as a trend
indicator. Besides, upon gradual increase in the packet generation rate $R$,
the first node to be congested may no longer be the one whose degree is
$k_{\min}$ or $k_{\max}$.
Nevertheless, the maximum point on the $\alpha - R_c$ curve due to the change
of the kind of node that is congested first is robust and generic as it is
stable upon small change in $n(k)$.
Of course, the expression for $R_c$ in Eq.~(\ref{Rc_full}) and the value of
$\alpha_c$ will be affected as a consequence of the break down of
Eqs.~(\ref{n_s_k_i_form}) and~(\ref{dir_dis1}).
Fortunately, as $\mu$ and $\nu$ are independent of $N$ and are not
sensitively dependent on $\alpha$, we conclude that $\Xi$ is almost $N$
independent in the large $N$ limit. More importantly, within about 10\%
accuracy, we may regard $\Xi$ as independent of $\alpha$.
Thus, the general trend of $R_c$ expressed in
Eqs.~(\ref{Rc_limit1})--(\ref{Rc_limit3}) is still valid.
That is to say, for sufficiently large $N$, $R_c$ is approximately proportional
to $1/(\gamma - \alpha - 2) \equiv 1/(1 - \alpha)$ for $\alpha < \alpha_c$. And
the proportionality constant is independent of $N$.
Besides, $R_c \sim 1/ [(1-\alpha) N^{(\alpha+1)/2}]$ for $-1 < \alpha < 1$
and $R_c \sim 1 / [N (\alpha - 1)]$ for $\alpha > 1$.
We are going to test these predictions using large scale numerical simulations
in Sec.~\ref{sec:sim}.

\subsection{Implications To The PNNN+PIA Model \label{subsec:with_PIA}}

\begin{figure}[t]
\centering
\includegraphics[width=\columnwidth]{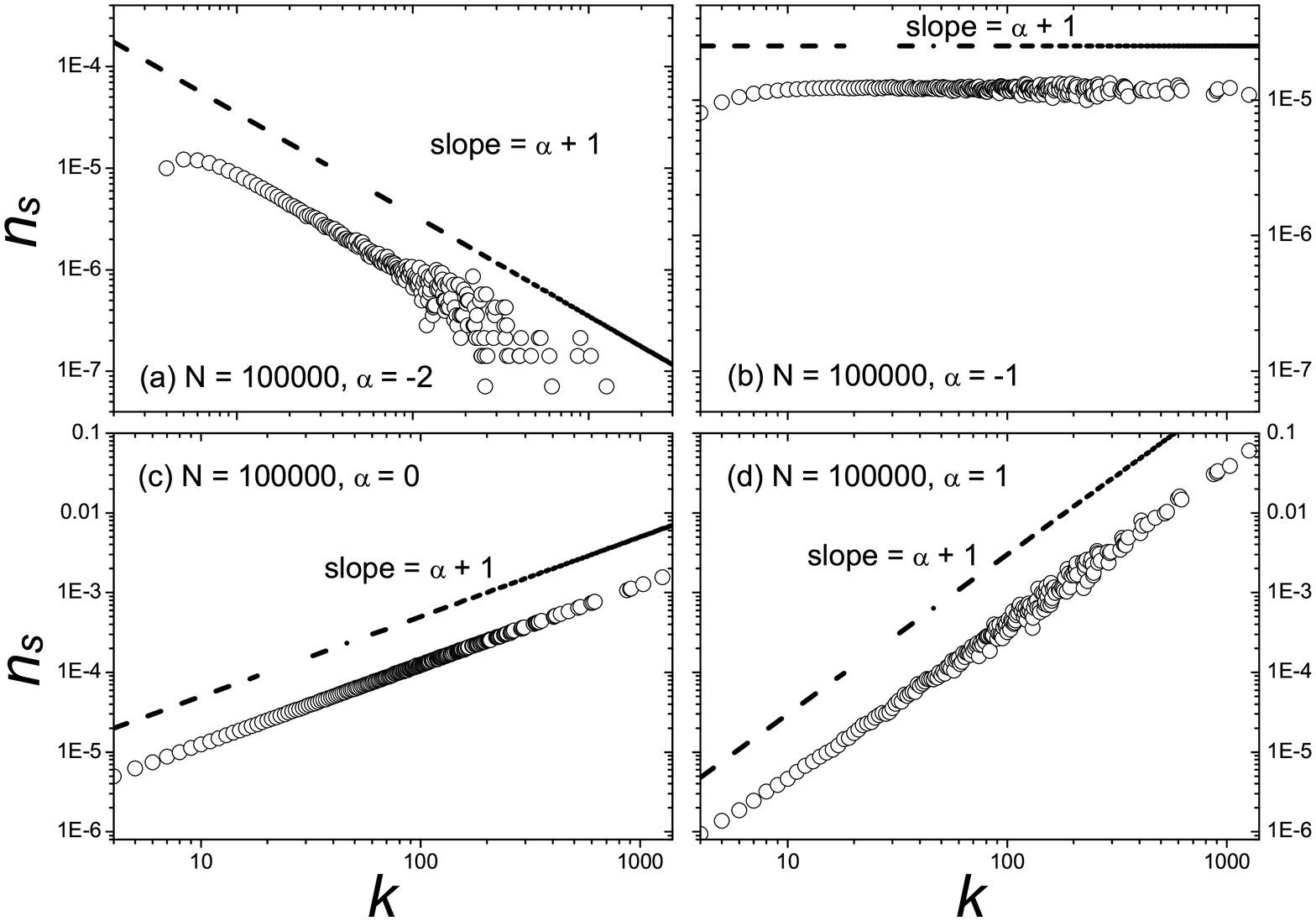}
\caption{\label{F:dsp-PIA} Log-log plot of the distribution of number of DSPs
 $n_s$ against the degree of node $k$ in black dots for PNNN-PIA with $m = 4$,
 $C = 1$ and $R = 5$ for network size $N = 10^5$ and preferential delivering
 exponents $\alpha$. The dashed line with slope $\alpha+1$ in each subplot is
 drawn for comparison purpose.}
\end{figure}

\begin{figure}[t]
\centering
\includegraphics[width=\columnwidth]{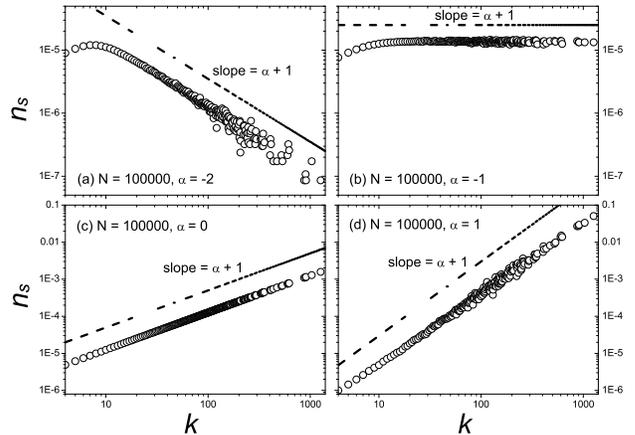}
\caption{\label{F:dsp+PIA} Log-log plot of the distribution of number of
 DSPs $n_s$ against the degree of node $k$ for PNNN+PIA. All parameters used
 are the same as those in Fig.~\ref{F:dsp-PIA}.}
\end{figure}

\subsubsection{Beyond The Simplifying Assumption \label{subsubsec:real_case}}
In reality, Eqs.~(\ref{n_s_k_i_form}) and~(\ref{dir_dis1}) are not exact.
Although it is much harder to modify the mean-field analysis in
Sec.~\ref{subsec:no_PIA} to take the PIA rule into account, we can still argued
the behavior of the PNNN+PIA model qualitatively. First, we consider the effect
of PIA rule on $n_s(k)$. Clearly, PIA rule makes $\Pi_{ji}$ in
Eq.~(\ref{free_eqn_t}) historical path dependent. Thus, we can no longer apply
the trick in Eq.~(\ref{no_assort}) to give a simple expression for $n_s (k)$.
Nevertheless, we may argue the behavior of $n_s (k)$ as follows.
In the case of $\alpha < 0$,
Eq.~(\ref{free_prob}) implies that packets are preferentially being forwarded
to small degree nodes.
However, the PIA rule forbids a packet to travel through the same edge
more than twice. Therefore, compared with the situation without the PIA rule, a
packet is less likely to be forwarded to a small degree node on average.
On the other hand, the PIA rule has relatively little effect on high degree
nodes. This is because of two reasons: first, packets are less likely to
travel to these nodes; and second, packets located at these nodes generally
have a large number of possible nodes to be forwarded to in the next time step.
Thus, we expect that the same scaling behavior for $n_s(k)$
found in Eq.~(\ref{n_s_k_i_form}) is observed in the presence of PIA rule.
Nonetheless, the domain of $k$ in which this scaling law holds is reduced as
the value of the lower cutoff of the scaling law increases as a consequence of
the PIA rule.
Furthermore, below this lower cutoff point, the value of $n_s(k)$ is smaller
that the case when the PIA rule is not adopted.

\begin{figure}[t]
\centering
\includegraphics[width=\columnwidth]{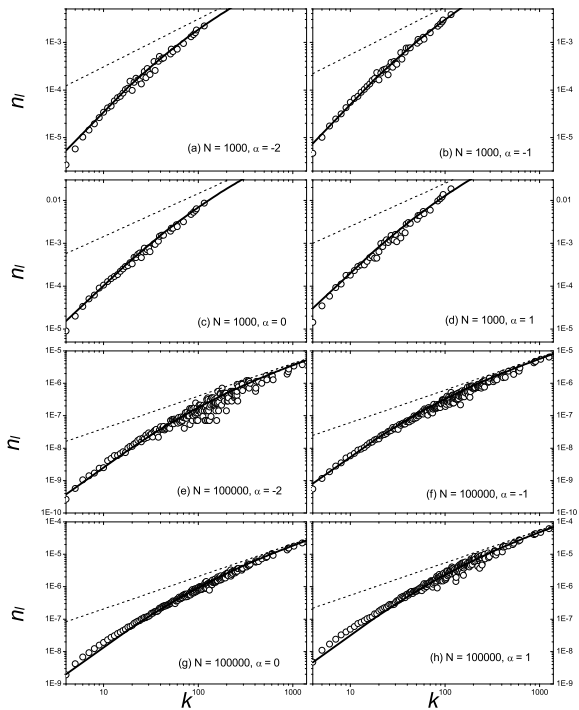}
\caption{\label{F:dlp-PIA} Log-log plot of the distribution of number of DLPs
 $n_l$ against degree of nodes $k$ for PNNN-PIA. The solid curve in each
 subplot is the prediction according to Eq.~(\ref{dir_dis1}) with $\mu$ and
 $\nu$ treated as free fitting parameters. And the dotted line in each subplot
 is the asymptote of the corresponding solid curve. All parameters used in the
 simulations are the same as those in Fig.~\ref{F:dsp-PIA}.}
\end{figure}	

\begin{figure}[t]
\centering
\includegraphics[width=\columnwidth]{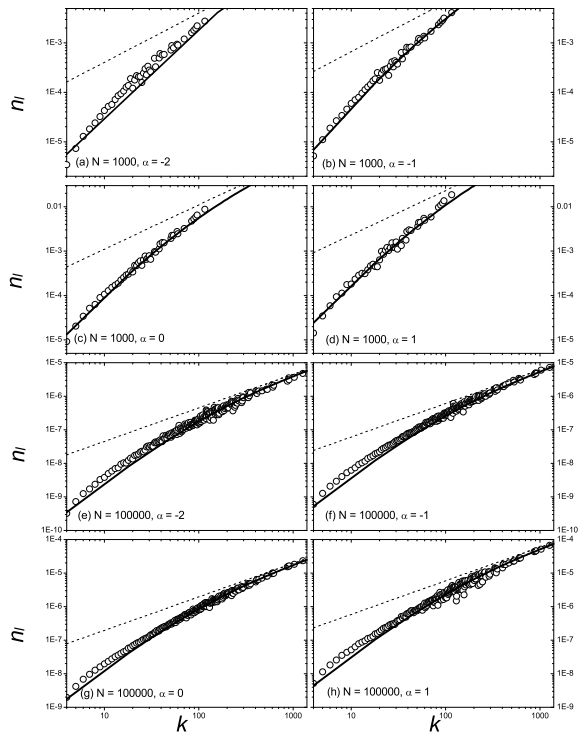}
\caption{\label{F:dlp+PIA} Log-log plot of the distribution of number of DLPs
 $n_l$ against degree of nodes $k$ for PNNN+PIA. The detailed procedure is
 adapted from the descriptions in Fig.~\ref{F:dlp-PIA}.}
\end{figure}

Applying similar arguments in the previous paragraph to the case of $\alpha >
0$, we conclude that it is more likely to
forward a packet between two large degree nodes. Since the number of nodes with
degree $k$ decreases as $k$ increases, the combination of
Eq.~(\ref{free_prob}) and the PIA rule will decrease (increase) the value of
$n_s(k)$ in Eq.~(\ref{n_s_k_i_form}) for
$k\lesssim k_{\max}$ ($k \ll k_{\max}$ and $k \gg k_{\min}$).
Therefore, the domain in which the scaling behavior of
Eq.~(\ref{n_s_k_i_form}) holds only for $k \ll k_{\max}$ and
$k \gg k_{\min}$.
To summarize, we have argued the validity of
Eq.~(\ref{n_s_k_i_form}) in the large $N$ limit for the PNNN+PIA model over a
reduced domain of $k$.
In addition, the value of
$\langle n_s(k_i) \rangle_{i\in {\mathbb V}}$ decreases in the presence of
PIA rule.

\begin{figure}[t]
\centering
\includegraphics[width=\columnwidth]{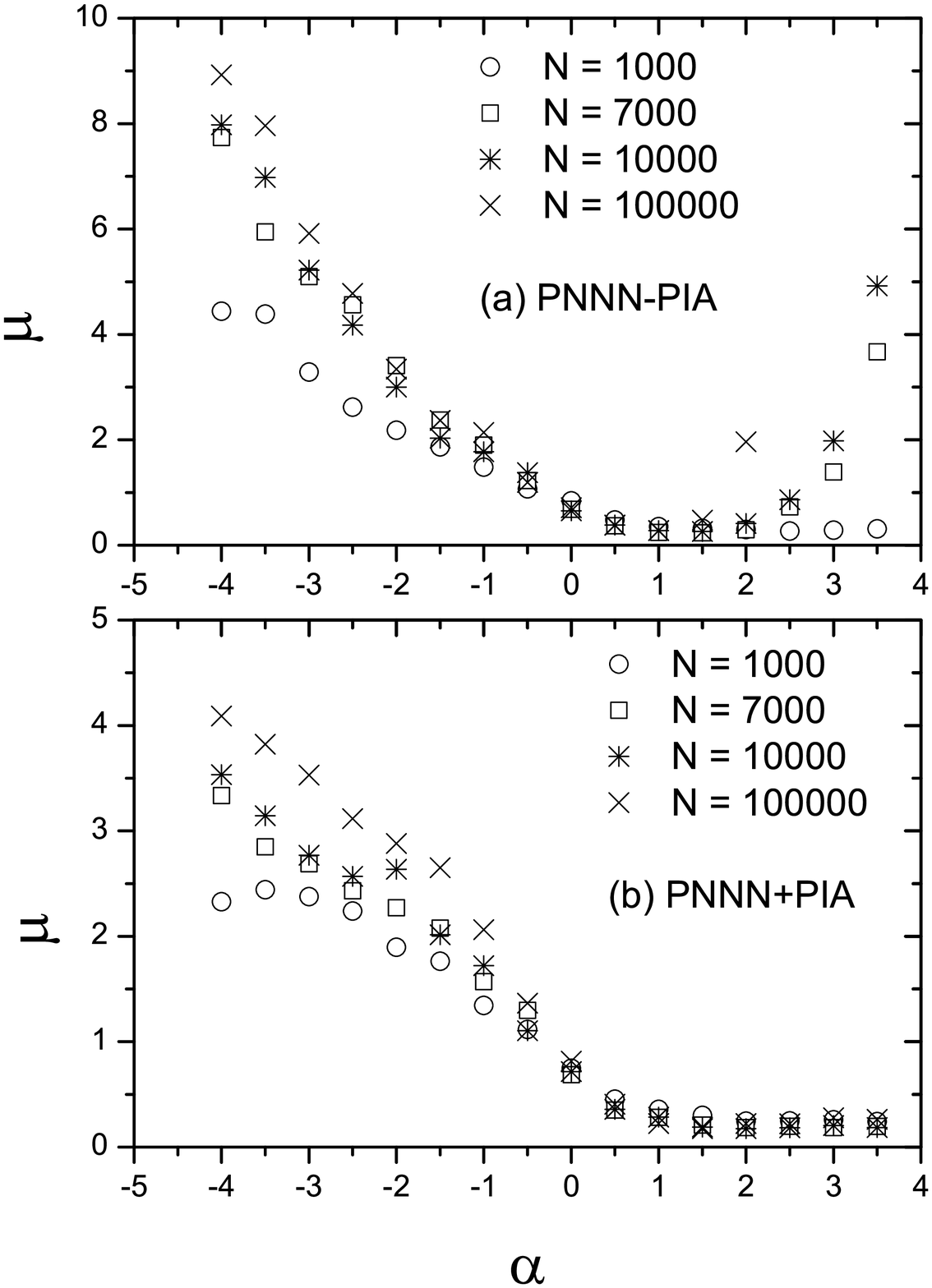}
\caption{\label{F:mu_graph} Plots of $\mu$ against $\alpha$ for (a)~PNNN-PIA
 and (b)~PNNN+PIA for various values of $N$.}
\end{figure}

\begin{figure}[t]
\centering
\includegraphics[width=\columnwidth]{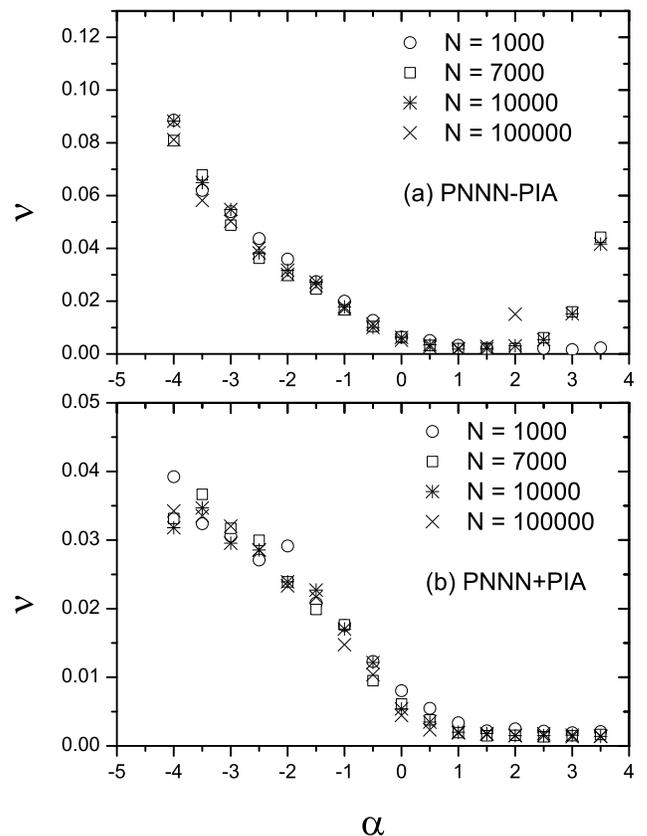}
\caption{\label{F:nu_graph} Plots of $\nu$ against $\alpha$ for (a)~PNNN-PIA
 and (b)~PNNN+PIA for various values of $N$.}
\end{figure}

How about the effect of PIA rule on $n_l(k)$? The PIA rule surely reduces
both $\mu$ and $\nu$ by forbidding excessive routing through the same edge.
Besides, the value of $\langle n_s (k_i)\rangle_{i\in {\mathbb V}}$ is also
reduced. But interestingly, unlike Eq.~(\ref{free_eqn_t}), the presence of
PIA rule in no way affects the functional form of Eq.~(\ref{dir_dis1}) as
the derivation of Eq.~(\ref{dir_eqn_t}) is also valid in this case. This is
because the PIA rule cannot prevent a DLP from reaching its destination
unless the distance between the destination and the initial generation
point of the packet is less than two. (This is because at the first
instance when a DSP is forwarded to a node $i$ with distance two from the
packet destination. The DSP will turn into a DLP in the next time step. 
More importantly, this message packet must never pass through any shortest
path connecting node $i$ and the packet destination. Hence, the PIA rule
does not prevent this packet from moving along this shortest path.) And the
probability for such case is negligible in the large $N$ limit. Note that
$n_l(k)$ is more seriously affected by
$\langle n_s (k_i)\rangle_{i\in {\mathbb V}}$ than by $\mu$ or $\nu$. So,
we expect that $n_l(k)$ decreases with the introduction of PIA rule. But
its percentage decrease is not as large as that of $n_s (k_{\max})$.

We now move on to study the effect of PIA rule on the values of $\alpha_c$ and
$R_c$. Recall that without PIA rule, $\alpha_c = -1$. Let us consider the case
of $\alpha > -1$ first. In this case, both $n_s(k)$ and
$n_l(k)$ are increasing functions of $k$ for $\alpha > -1$ with or without
PIA rule. So, upon a gradual increase in the packet injection rate, the first
node to congest must be the one with a large degree. From the arguments in this
Subsection, we know that for $k\approx k_{\max}$, $n(k) \equiv n_s(k) + n_l(k)$
decreases with the introduction of PIA rule. Hence, the critical packet
injection rate $R_c$ increases with the introduction of PIA rule. Certainly,
the percentage increase in $R_c$ depends on the values of $N$, $m$ and
$\alpha$ used; and the above arguments in no way imply that the percentage
change is huge. Indeed, it is quite possible that the increase in $R_c$ is
negligible in some cases.

In contrast, when $\alpha < -1$, Eq.~(\ref{dir_dis}) together with 
the arguments in this Subsection tell us that $n (k_{\min}) \approx
n_s(k_{\min})$ decreases more
rapidly than $n(k_{\max}) \approx n_l(k_{\max}) \sim
\left\langle n_s(k_i)\right\rangle_{i\in {\mathbb V}} k_{\max}$ with
the introduction of PIA rule.
Consequently, $R_c$ increases in the presence of PIA rule.
More importantly, for a finite $N$, one may find an $\alpha$ slightly less
than $-1$ such that $n(k_{\min}) < n(k_{\max})$. In other words, a large
degree instead of a small degree node gets congested at $R_c$ for this
value of $\alpha$. Therefore, we conclude that
$\alpha_c$ decreases and $R_c$ increases in the presence of PIA rule.
Note that once again the decrease in $\alpha_c$ may be insignificant
in some cases.

Finally, we expect that the general trend of $R_c$ for PNNN-PIA
described in Sec.~\ref{subsubsec:real_case} also applies to PNNN+PIA.
Obviously, our predictions are different from the numerical results of Yin
\emph{et al.} reported in Ref.~\cite{Yin}, which claimed that $R_{c}$ was a
decreasing function of $\alpha$ for PNNN+PIA.
In Sec.~\ref{sec:sim}, we show that this is
partly due to the fact that the network size $N$ used in their simulation is
not large enough so that finite-size effect seriously affects their
conclusions.

\section{Comparison with our numerical simulations \label{sec:sim}}

We want to check the validity of
our mean-field analysis reported in the previous Section as well as to
to understand the origin of the discrepancy between
our present work and the numerical results obtained by Yin \emph{et al.}
in Ref.~\cite{Yin}.
And we do so by performing numerical simulations using
larger values of $N$. Moreover, unlike Ref.~\cite{Yin}, we
allow $R$ to take on non-integer values.

\begin{figure}[t]
\centering
\includegraphics[width=\columnwidth]{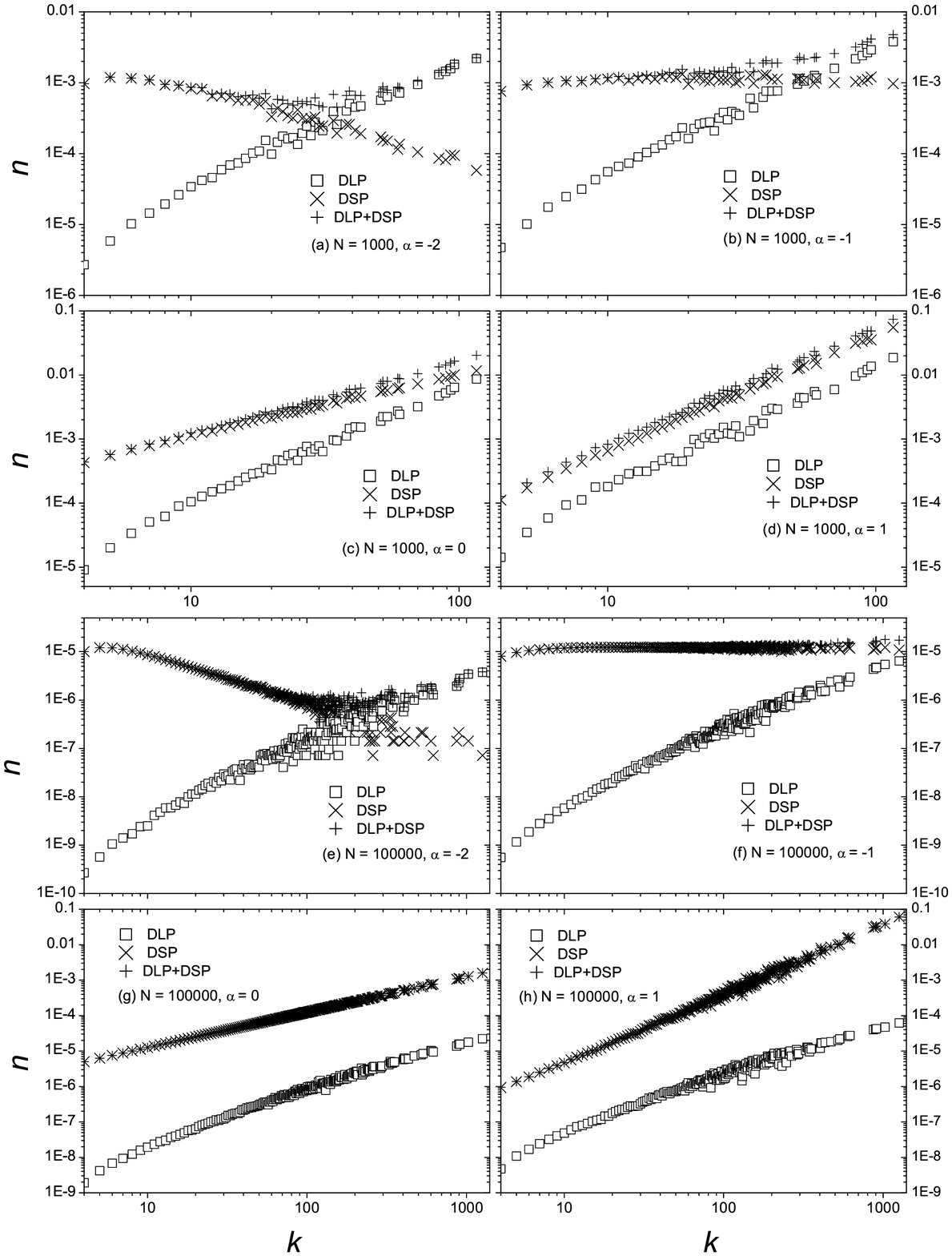}
\caption{\label{F:packet-PIA} The average number of packets $n$ against degree
 of nodes $k$ for PNNN-PIA at $R = R_c$. All parameters used in the simulations
 are the same as those in Fig.~\ref{F:dsp-PIA}.}
\end{figure}	

\begin{figure}[t]
\centering
\includegraphics[width=\columnwidth]{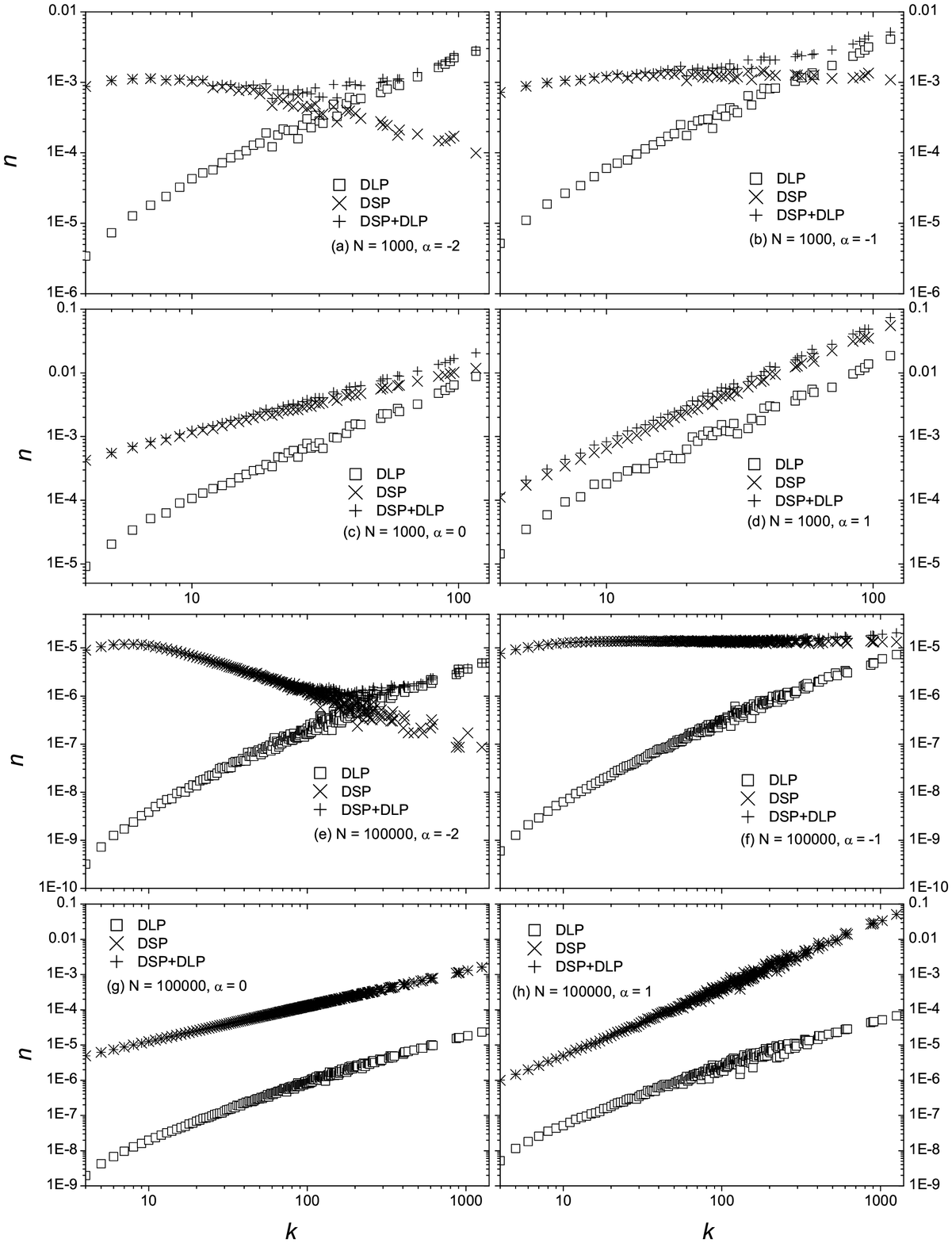}
\caption{\label{F:packet+PIA} The average number of packets $n$ against degree
 of nodes $k$ for PNNN+PIA at $R = R_c$. All parameters used in the simulations
 are the same as those in Fig.~\ref{F:dsp-PIA}.}
\end{figure}  

Perhaps one of the reasons why Yin \emph{et al.} reported numerical
simulations of PNNN+PIA up to $N=5000$ only~\cite{Yin} is that a lot
of memory is needed to store the message packets present in the network as
well as their historical paths. In fact, this straight-forward numerical
simulation method is not practical for $N\gtrsim 10000$.
Here we introduce a much less memory intensive way to numerically find $R_c$.
Observe that the connectedness of BA network and the message forwarding
rules of PNNN$\pm$PIA make the message packets in PNNN$\pm$PIA ergodic.
Also, recall from Sec.~\ref{sec:model} that message packets behave like
independent
particles as long as there are no more than $C$ packets staying in a node at
any time.
Although occasionally more than $C$ message packets may be present
in a node in the free-flow phase, by ergodicity we expect that the statistical
properties of PNNN$\pm$PIA below the critical packet injection rate $R_c$ can
still be simulated by regarding each message packet as independent particle
throughout.
Therefore, the statistical behavior of PNNN$\pm$PIA for $R < R_c$ can
be found as follows: We first numerically simulate the 
ensemble-averaged time evolution of a particular free-flow phase situation in
which there is exactly one message packet in the network at all times.
By ergodicity, the ensemble-averaged number of packet present in a node
obtained in the above simulation equals the (time-averaged) number of packet
in that node when the packet
injection rate $R$ is $1 / \langle\tau\rangle$, where $\langle\tau\rangle$
denotes the mean packet lifetime.
(This choice of $R$ does not contradict with the prediction of
Eqs.~(\ref{Rc_limit2}) and~(\ref{Rc_limit3}) that $R_c \rightarrow 0$ in the
limit of large $N$ whenever $\alpha > \alpha_c$. This is because the mean
packet lifetime $\langle\tau\rangle$ scales like $N^\beta$ with $\beta \geq
2$.)
Below the critical packet injection rate $R_c$, the distributions
$n_s(k)$, $n_l(k)$ and $n(k)$ are directly proportional to the packet injection
rate $R$. Consequently, $R_c$ is equal to $C / \langle\tau\rangle \max_k n(k)$
where $\max_k n(k)$ is the maximum value of $n(k)$ over all $k$ for
the case of $R = 1/\langle\tau\rangle$. Clearly, this method can compute
$R_c$ accurately and efficiently.
As only one message packet is used at any time in the simulation,
this method requires much less memory than the straight-forward numerical
simulation approach.
We further verify the validity of this ensemble-averaged simulation method by
successfully reproducing the numerical simulation results reported by
Yin \emph{et al.} in Ref.~\cite{Yin} (modulo the fact that they restricted
$R$ to integers). (Actually, the value of $m$ used for their PNNN-PIA
simulation is 4 instead of 5~\cite{Yin_private}.)
Therefore, we adopt this new method in our subsequent numerical studies.

While the simulations of Yin \emph{et al.} in Ref.~\cite{Yin} was performed
in for $\alpha\in [-4,2]$, ours is done in a slightly large parameter range of
$[-4,4]$. Actually, we find that
the bias in forwarding a DSP according to Eq.~(\ref{free_prob}) for
$|\alpha|$ close to $4$ is already so high that
the data obtained from our simulations are no longer very reliable.
And reliable results for $|\alpha| \gtrsim 4$ has to be obtained by much
longer simulation time with the aid of a higher precision pseudo random
number generator.

\begin{figure}[t]
\centering
\includegraphics[width=\columnwidth]{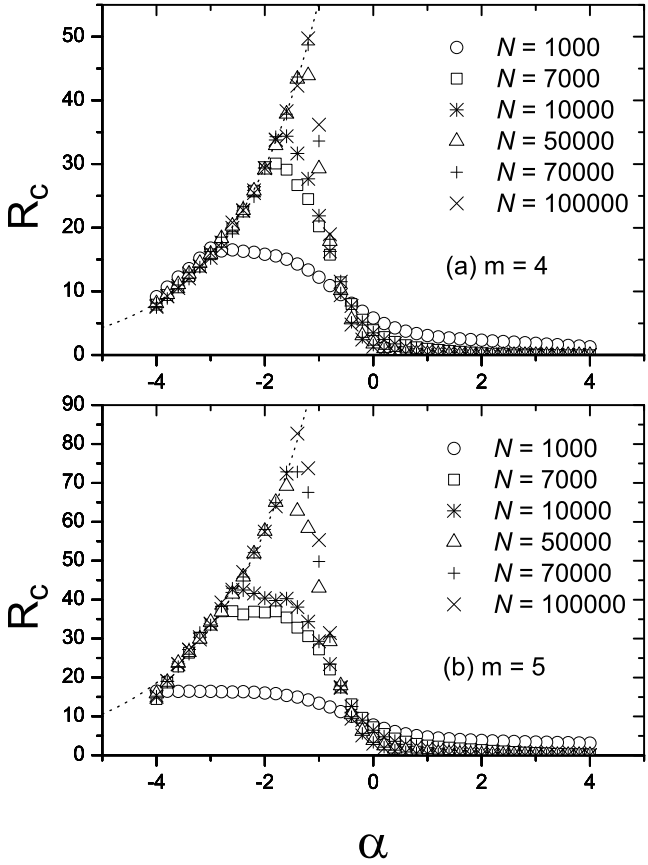}
\caption{\label{F:Rcvsalpha-PIA} The $R_c$ against $\alpha$ curve for PNNN-PIA
 with $C = 1$ and (a)~$m = 4$, and (b)~$m = 5$. The dashed curve in each
 subplot is our mean field analytical prediction based on
 Eq.~(\ref{Rc_limit1}). More precisely, the dashed curve is the best fit curve
 obtained from Eq.~(\ref{Rc_limit1}) by treating $\Xi$ as a free parameter
 independent of $\alpha$.}
\end{figure}

\begin{figure}[t]
\centering
\includegraphics[width=\columnwidth]{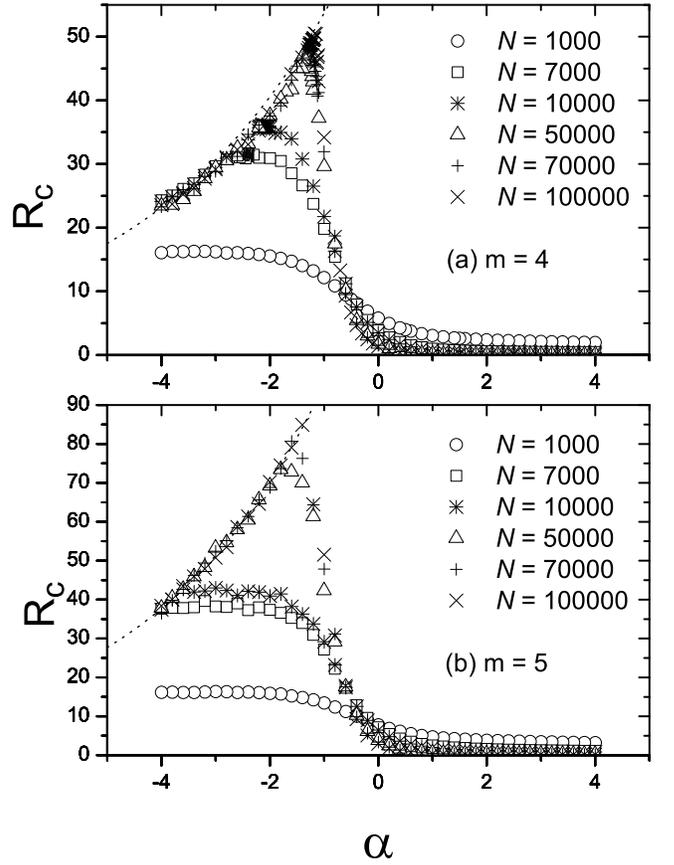}
\caption{\label{F:Rcvsalpha+PIA} The $R_c$ against $\alpha$ curve for PNNN+PIA.
 Parameters used are the same as those in Fig.~\ref{F:Rcvsalpha-PIA}.}
\end{figure}

Let us begin by checking the validity of our assumptions made in
Sec.~\ref{sec:cal}. Figs.~\ref{F:dsp-PIA} and~\ref{F:dsp+PIA} show typical
$n_{s}(k)$ curves obtained from our numerical simulations of PNNN-PIA and
PNNN+PIA, respectively.
They show that $n_{s}(k)$ indeed follows a power law with exponent $\alpha+1$
over most of the parameter range for PNNN$\pm$PIA for sufficiently large $N$.
Furthermore, the domain of validity of the power law is reduced with
the introduction of PIA rule.
More importantly, in the case of PNNN+PIA, the ways how $n_{s}(k)$ deviates
from the power law for small and large $k$ are consistent with our predictions
in Sec.~\ref{subsec:with_PIA}. That is to say, $n_s(k)$ is less (greater) than
the value obtained by Eq.~(\ref{n_s_k_i_form}) for $k \approx k_{\min}$
($k \approx k_{\max}$). In this respect, our assumption of ignoring
degree-degree correlation between neighboring nodes in obtaining $n_s(k)$
is not bad.

Next, we examine the validity of Eq.~(\ref{dir_dis1}) for PNNN$\pm$PIA.
Figs.~\ref{F:dlp-PIA} and~\ref{F:dlp+PIA} plot $n_l$ as a function of $k$
obtained from our simulation of PNNN-PIA and PNNN+PIA, respectively.
Our simulation results for $n_l(k)$ agree quite well with the solid curves,
namely, our mean field prediction given by Eq.~(\ref{dir_dis1}).
The dotted lines in Figs.~\ref{F:dlp-PIA} and~\ref{F:dlp+PIA}
show the asymptotic behavior of the solid curve in the limit of large $k$.
By comparing our simulated data points with the dotted lines, we find that
for $N$ as small as $1000$, $n_l(k)$ does not reach the linear scaling
regime at all.
And for $N = 100000$, $n_l(k)$ attains linear scaling for $k
\gtrsim 200$. In fact, we discover from our
simulation that $n_l(k)$ scales like a linear function of $k$ around
$k\lesssim k_{\max}$ only when $N\gtrsim 10000$.

As shown in Figs.~\ref{F:mu_graph} and~\ref{F:nu_graph},
$\mu$ and $\nu$ are independent of $N$ for PNNN$\pm$PIA provided that
$N\gtrsim 7000$ and $|\alpha| \lesssim 3$.
We believe that the discrepancy for $\mu$ when $N = 1000$ in
Fig.~\ref{F:mu_graph} is the result of finite size effect. And as we have
already discussed earlier in this Section, we think that the
discrepancies for $\mu$ and $\nu$ for $|\alpha| \gtrsim 3$ are due to the
limitations of our simulation time and pseudo random number generator used.
In any case, Figs.~\ref{F:mu_graph} and~\ref{F:nu_graph} verify that
$\mu$ and $\nu$ are not sensitively dependent on $\alpha$.
In fact, $\mu$ and $\nu$ are of order of $1$ and $0.01$
respectively over most of the range of $\alpha$ we have studied. And
in line with our expectation, $\mu$ and $\nu$ decrease with the introduction
of PIA rule.

Figs.~\ref{F:packet-PIA} and~\ref{F:packet+PIA} depict the general trend of
$n_s(k)$, $n_l(k)$ and $n(k) = n_s(k) + n_l(k)$ near $R = R_c$ for
PNNN-PIA and PNNN+PIA, respectively.
They show that for a sufficiently small $\alpha$, the degree of
the congested node at $R = R_c$ is generally close but not equal to $k_{\min}$.
This is not surprising because there are numerous nodes with degree close to
$m$. Local conditions such as the degrees of the neighbors of these small
degree nodes can vary a lot. Combined with the break down of the scaling
relation in Eq.~(\ref{n_s_k_i_form}), jamming may occur at a node whose
degree is slightly greater than $k_{\min}$ when $R = R_c$.
In contrast, Figs.~\ref{F:packet-PIA} and~\ref{F:packet+PIA} show that for
a sufficiently large $\alpha$, jamming almost always occurs in the
highest degree node in the network. This is because for
a generic BA network with a large but fixed $N$, there is a considerable
difference between the degree of the most connected and second most connected
nodes. Thus, $n_l$ for the most connected node is almost surely greater than
that for the slightly less connected ones.
Most importantly, our simulations find that the transition between these two
types of congested nodes at $R = R_c$ occurs
at a rather well-defined critical $\alpha_c$ for $N\gtrsim 1000$. And
the value of $\alpha_c$ depends on the value of $N$ as well as on whether
the PIA rule is adopted or not.

\begin{figure}[t]
\centering
\includegraphics[width=\columnwidth]{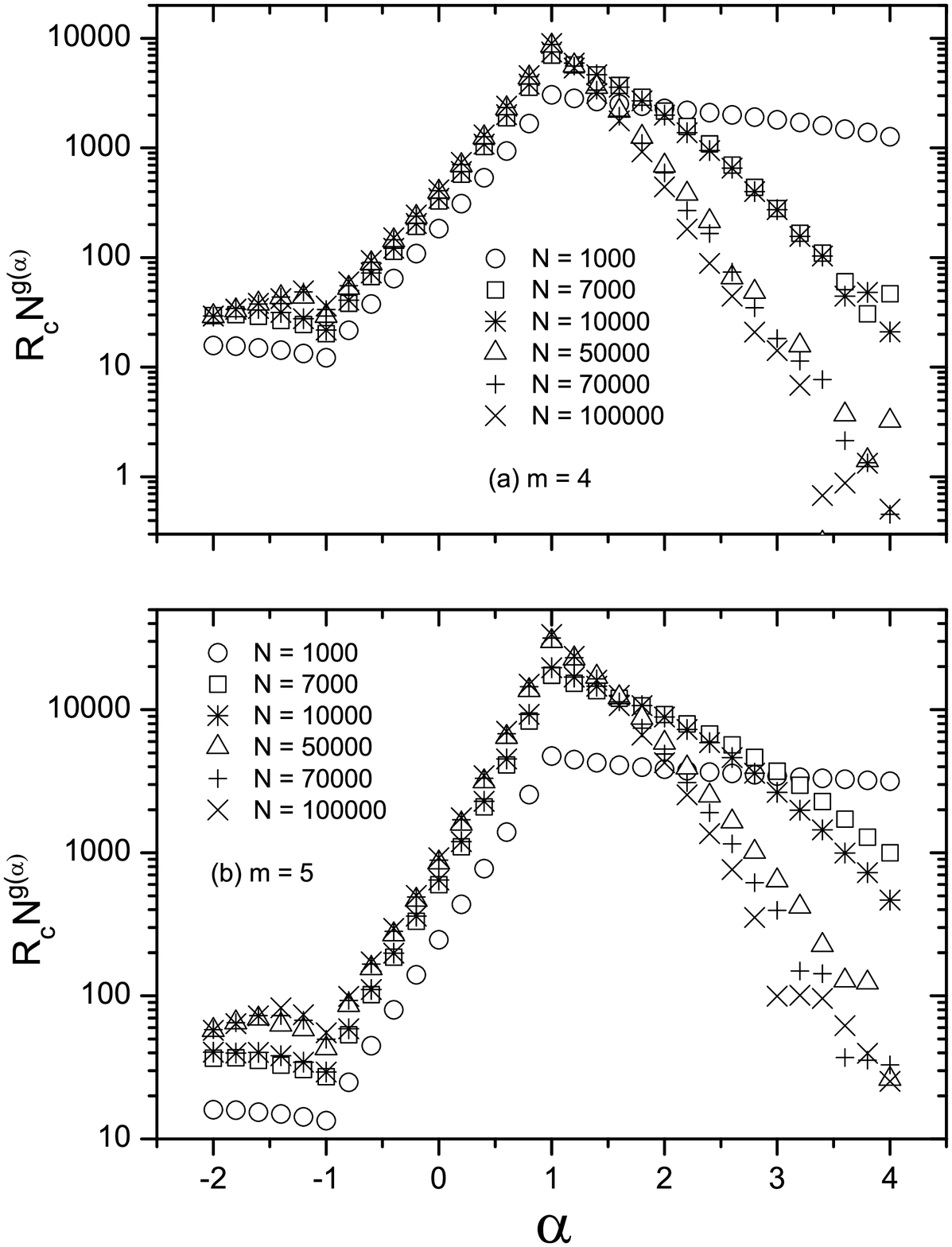}
\caption{\label{F:Rc_scaling-PIA} The $R_c N^{g(\alpha)}$ against $\alpha$
 curve for PNNN-PIA. Parameters used are the same as those in
 Fig.~\ref{F:Rcvsalpha-PIA}.}
\end{figure}

\begin{figure}[t]
\centering
\includegraphics[width=\columnwidth]{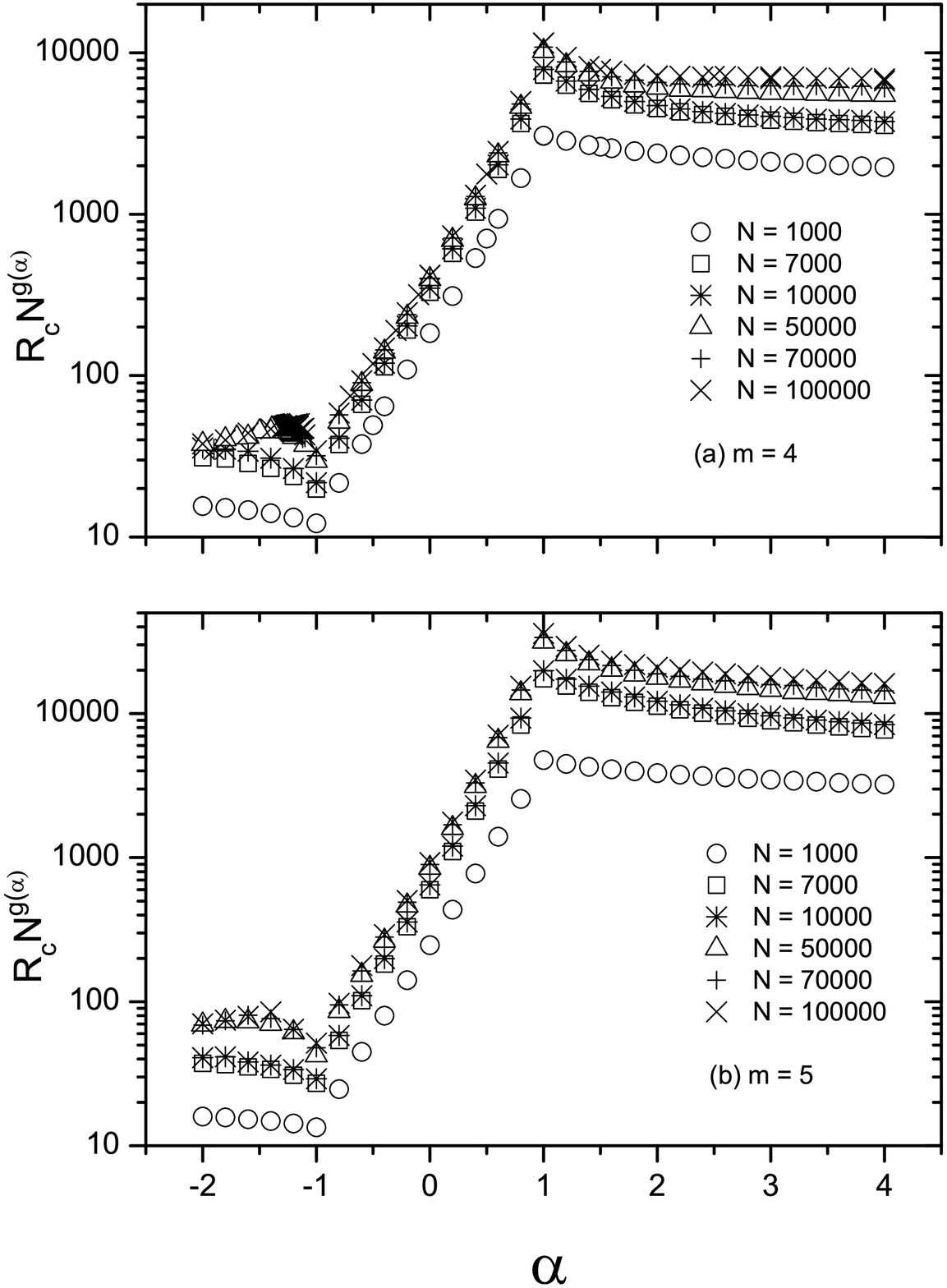}
\caption{\label{F:Rc_scaling+PIA} The $R_c N^{g(\alpha)}$ against $\alpha$
 curve for PNNN+PIA. Parameters used are the same as those in
 Fig.~\ref{F:Rcvsalpha-PIA}.}
\end{figure}

After finish justifying the validity of the approximations made in our mean
field analysis, we now move on to compare our mean field calculations and
numerical simulation results with the numerical findings of Yin \emph{et al.}
reported in Ref.~\cite{Yin}.
As the $R_c$ against $\alpha$ curves in Figs.~\ref{F:Rcvsalpha-PIA}
and~\ref{F:Rcvsalpha+PIA} shown, the general trend of $R_c$ we find in our
numerical simulations agrees quite well with the predictions of our mean
field theory for both PNNN-PIA and PNNN+PIA. In particular, we discover that
for fixed $N$ and $m$, $R_c$ is an increasing (decreasing) function of
$\alpha$ for $\alpha < \alpha_c$ ($\alpha > \alpha_c)$.
Besides, $\alpha_c$ decreases and $R_c$ increases with the introduction of
PIA rule although the change is not significant for large $N$ and small $m$.
Recall from Eq.~(\ref{Rc_limit1}) and the discussions in
Sec.~\ref{subsubsec:real_case} and~\ref{subsec:with_PIA} that in the
large $N$ limit, $\Xi$ should be roughly a constant over the parameter range of
interest and $R_c$ should
roughly scales like $1/(1-\alpha)$ whenever $\alpha < \alpha_c$.
This is exactly what we find in Figs.~\ref{F:Rcvsalpha-PIA}
and~\ref{F:Rcvsalpha+PIA}.
More generally, Eqs.~(\ref{Rc_limit1})--(\ref{Rc_limit3}) imply that
$R_c N^{g(\alpha)}$ should be $N$ independent, where
\begin{equation}
 g(\alpha) = \left\{ \begin{array}{ll}
 0 & \enspace \mbox{for} \enspace \alpha < -1 , \\
 (\alpha+1)/2 & \enspace \mbox{for} \enspace -1 < \alpha < 1 , \\
 1 & \enspace \mbox{for} \enspace \alpha > 1 .
 \end{array} \right. \label{g_def}
\end{equation}
As shown in Figs.~\ref{F:Rc_scaling-PIA} and~\ref{F:Rc_scaling+PIA},
$R_c N^{g(\alpha)}$ is indeed $N$ independent for
$\alpha < 1$ ($\alpha > 1$) provided that $N\gtrsim 10000$
($N\gtrsim 500000$). Again, the discrepancy for $\alpha > 3$ is probably caused
by insufficient sampling and the finite precision of our pseudo random
number generator.

As for the critical preferential delivering exponent $\alpha_c$, we find that
it decreases as $m$ increases for a fixed $N$. This can be explained as
follows: Recall that the number of 4-cycles in a BA network scales like
$\left[ m \log (N) /2 \right]^4 / 4$~\cite{BA_4-cycle}.
So, by increasing $m$ while fixing $N$, the proportion of 4-cycles in the
network increases. In other words, the assumption of neglecting the
effect of 4-cycles in our mean-field analysis
reported Sec.~\ref{sec:cal} becomes less valid. By going through the
analysis in Sec.~\ref{sec:cal} once more, it is not difficult to see that
although the scaling relations in Eqs.~(\ref{n_s_k_i_form})
and~(\ref{dir_dis}) are robust against the presence of 4-cycles,
the $(k-1)$ factor in Eq.~(\ref{dir_dis1}) should be replaced by
$(k-\zeta)$ for some $\zeta > 1$. This change decreases the value of $n_l(k)$
for a fixed $N$, therefore making the small degree node harder to jam. This is
the reason why the presence of large number of 4-cycles reduces the value of
$\alpha_c$. 

In the case of $m=4$, Figs.~\ref{F:Rcvsalpha-PIA}(a)
and~\ref{F:Rcvsalpha+PIA}(a) show that $\lim_{N\rightarrow\infty} \alpha_c$ is
very close to $-1$ for PNNN$\pm$PIA. Combined with the validity of
Eqs.~(\ref{Rc_limit1}) and~(\ref{Rc_limit2}) as depicted in
Figs.~\ref{F:Rc_scaling-PIA}(a) and~\ref{F:Rc_scaling+PIA}, we conclude that
$R_c$ approaches to its maximum value at $\alpha_c = -1$ in the large $N$ limit.
In contrast, for simulation up to $N = 100000$,
$\alpha_c$ does not seem to converge to $-1$ in the case of $m = 5$. As we have
discussed in the last paragraph, we believe that this is due to the existence
of large amount of 4-cycles. Since for $m = 5$, the number of 4-cycles is less
than about $N/10$ provided that $N\gtrsim 10^7$, we believe that $\alpha_c$
should converge to $-1$ by using networks at least about 100~times larger than
our currently used ones. Unfortunately, such simulation is beyond the
current computing capacity of our group.

Now, let us compare our findings with that of Yin \emph{et al.}'s in
Ref.~\cite{Yin}. Fig.~\ref{F:Rcvsalpha+PIA} show that the
simulations performed on a $N = 1000$ network does not reveal the
thermodynamic behavior of the system due to serious
finite size corrections. Actually, if they had extended their numerical
simulations to $\alpha$ as small as about $-8$ (which unfortunately requires
much longer computational time and the use of a high precision pseudo
random number generator), they should have revealed the maximum point on the
$\alpha-R_c$ curve, thereby discovering the critical $\alpha_c$.

\section{Discussions \label{sec:sum}}

To summarize, we have pointed out that the PNNN+PIA model is not a good model
of network traffic due to the hidden communication cost involved.
In addition, we have carefully performed a mean-field analysis of the message
packet dynamics for a network traffic model with PNNN routing strategy on BA
network with or without PIA by Yin \emph{et al.} in Ref.~\cite{Yin}. The main
feature of our analysis is that we divide the message packets into two groups,
namely, the DSPs and DLPs. To check the validity of our mean-field results, we
introduce a new method to simulate the critical packet injection rate $R_c$
that requires much less memory. This enable us to carry out an extensive
numerical simulation to study the so-called $\alpha-R_{c}$ curve for larger
network size $N$ with the message packet injection rate $R$ taking on real
rather than integer values.

For a fixed finite network size $N$,
we discover that the $\alpha-R_c$ curve is in fact increasing (decreasing) for
$\alpha < \alpha_c$ ($\alpha > \alpha_c$). And we are able to explain this
behavior by means of our mean-field analysis.
In fact, both our mean-field calculations and our numerical simulations show
that the critical message generation rate $R_{c}$ attains its maximum value at
$\alpha_c = -1$ for models both with and without PIA rule in the limit of large
$N$.
In this respect, the role of PIA rule has little effect on the $\alpha-R_c$
curve even though the value of $R_c$ is increased by introducing the PIA rule.
At the same time, Eq.~(\ref{Rc_limit1}) tells us that $R_c$ is independent of
$N$ in the limits of $N\rightarrow\infty$ and $\alpha\rightarrow -1^-$.
This means that the PNNN mechanism is not efficient
in handling large scale BA network traffic. In fact, this result agrees with
those of Sreenivasan \emph{et al.}~\cite{Sreenivasan} who showed that
$R_c \le O(\sqrt{N})$ for a BA network with any routing
strategy.

One may apply our analysis to consider the extension of PNNN$\pm$PIA to the
case in which more extended local information of the network such as the
third nearest neighbors is used to forward a packet. It is not too difficult
to argue that $n_s(k) \sim k^{\alpha+1}$ and $n_l(k) \sim k$ in the large $N$
limit for this kind of models. Thus, it appears that straight-forward
generalizations of the PNNN packet forwarding rule are also not efficient to
handle large scale BA network traffic in the sense that the resultant maximum
possible value of $R_c$ is independent of $N$.  One has to find other type of
strategies in order to approach the upper bound of $O(\sqrt{N})$ for $R_c$.

In addition to the functional dependence of $R_c$ on $\alpha$, it is
instructive to study the nature of the phase transition between the free-flow
and jamming phases in PNNN$\pm$PIA. Nonetheless, our mean field analysis and
the trick used in our extensive numerical simulations are for free-flow phase
only. Further work has to be done to investigate this problem.

\begin{acknowledgments}
We thank B.-H. Wang for bringing his group's work to our attention and for his
valuable discussions.
We also thank the Computer Center of HKU for their helpful support in providing
the use of the HPCPOWER system for performing part of the simulations reported
in this paper.
\end{acknowledgments}

\bibliography{en3.3}

\end{document}